\documentclass[superscriptaddress,twocolumn,aps,prl,amsmath,floatfix] {revtex4} 
\usepackage{graphicx}
\usepackage{bm}

\begin{document}

\title{Rubber friction on wet and dry road surfaces: the sealing effect}
\author{B.N.J. Persson}
\affiliation{IFF, FZ-J\"ulich, 52425 J\"ulich, Germany}
\author{U. Tartaglino}
\affiliation{IFF, FZ-J\"ulich, 52425 J\"ulich, Germany}
\affiliation{International School for Advanced Studies (SISSA),
Via Beirut 2, I-34014 Trieste, Italy}
\affiliation{INFM Democritos National Simulation Center, Trieste, Italy}
\author{O. Albohr}
\affiliation{Pirelli Deutschland AG, 64733 H\"ochst/Odenwald, Postfach 1120, Germany}
\author{E. Tosatti}
\affiliation{International School for Advanced Studies (SISSA),
Via Beirut 2, I-34014 Trieste, Italy}
\affiliation{INFM Democritos National Simulation Center, Trieste, Italy}
\affiliation{International Center for Theoretical Physics (ICTP),
P.O.Box 586, I-34014 Trieste, Italy}

\begin{abstract}
Rubber friction on wet rough substrates at low velocities is 
typically $20-30\%$ smaller than for the corresponding dry surfaces. 
We show that this cannot be due 
to hydrodynamics and propose a novel explanation based on a sealing 
effect exerted by rubber on substrate ``pools'' filled with water. 
Water effectively smoothens the substrate, reducing the major 
friction contribution due to induced viscoelastic deformations 
of the rubber by surface asperities. The theory is illustrated with 
applications related to tire-road friction.

\vskip 0.5cm \noindent
DOI: 10.1103/PhysRevB.71.035428 \hfill
Ref.: {\it Phys. Rev.} B {\bf 71}, 035428 (2005)
\vskip 0.5cm

\end{abstract}

\maketitle

{\bf 1. Introduction}

The study of sliding friction has attracted increasing interest 
during the last 
decade
thanks also to
the development of new experimental 
and theoretical approaches\cite{Persson,Meyer,Muser,Krim}. 
While some 
understanding has been gained about the origin and
qualitative properties of friction, first principle calculations
of friction forces (or friction coefficients) for realistic systems 
are in general impossible.
The basic reason for this is that friction usually is an 
interfacial property, often determined by the last few 
uncontrolled
monolayers of atoms or molecules at the interface. An extreme illustration
of this is diamond friction: the friction between two clean 
diamond surfaces in ultra high vacuum
is huge because of the strong 
interaction between the surface dangling bonds. However, when the
dangling bonds are saturated by monolayers of hydrogen atoms
(as they invariably are in real life conditions), 
friction becomes extremely low\cite{Filipse}.
Since most surfaces of practical use are covered by several monolayers of 
contamination molecules of unknown composition, the quantitative 
prediction of sliding friction coefficients
is generally impossible. An exception to this may be rubber friction on rough surfaces,
which is the topic of the present paper.
  
Rubber friction is a topic of extreme practical
importance, e.g., in the context of tires, wiper blades, conveyor belts
and sealings\cite{Moore}.
Rubber friction has several 
remarkable
properties. First, it may be huge, sometimes resulting in friction 
coefficients much higher than unity. Secondly,  
on very rough surfaces, e.g., in the context of a tire
sliding
on a road surface, it is mainly a {\it bulk} property of the rubber. 
That is, the substrate (or road)
asperities exert pulsating forces onto the rubber surface which, 
because of its high internal friction 
at the appropriate frequencies, results in a large dissipation of energy
in the rubber bulk (hysteresis contribution)\cite{PerssonJCP,Heinrich,PerssonTBP}. 
Finally, rubber friction 
is very sensitive to  temperature because of
the strong temperature dependence of the viscoelastic 
bulk properties of rubber-like materials.

Rubber friction on smooth substrates, e.g., a smooth glass 
surface, has two contributions, namely an adhesive (surface) and a hysteresis 
(bulk) contribution\cite{Moore,PerssonSS}. The adhesive contribution
results from the attractive binding forces between the rubber 
surface and the substrate.
These interactions are often dominated by weak van der 
Waals forces. However, because of the low elastic moduli 
of rubber-like materials, even when the applied
squeezing force is very 
gentle
this weak attraction may result in a nearly complete
contact between the solids at the interface
\cite{PerssonEur,Oliver},
resulting in
the large sliding friction force usually observed even for very smooth 
surfaces\cite{Grosch}.
The hysteresis
contribution results instead from the substrate 
roughness (even highly polished surfaces have
surface roughness, at least on the nanometer scale).

\begin{figure}[htb]
  \includegraphics[width=0.35\textwidth]{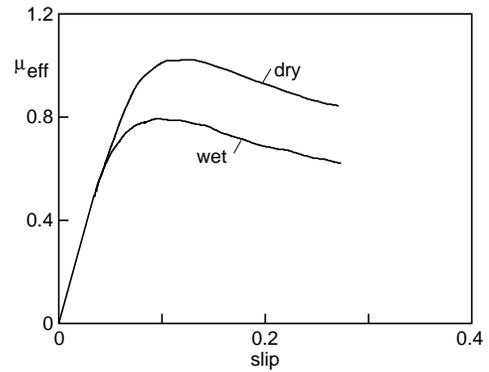}
  \caption{
  \label{measured.mu.slip}
A typical measured effective friction coefficient as a function 
of slip for dry and wet road surface.
See Sec. 3 for the definition of the slip.
    }
\end{figure}

For very rough surfaces the adhesive contribution to rubber 
friction will be much smaller 
than for smooth surfaces, mainly because of the small contact area. 
For a tire in contact with a road surface, for example, the actual contact 
area between the tire and the substrate is typically only 
$\sim 1 \%$ of the nominal footprint 
contact area\cite{PerssonJCP,Heinrich}. 
We have shown recently that the observed friction when a 
tire is sliding 
on a dry road surface can be calculated accurately 
by assuming it to be due entirely to internal damping in
the rubber (the hysteresis contribution)\cite{PerssonJCP,PerssonTBP}. 
This theory takes into account the pulsating
forces acting on the rubber surface from road asperities on many different
length scales, from the length scale $\lambda_0 \sim 1 \ {\rm cm}$, corresponding
to the largest road asperities, down to micro-asperities characterized by a wavelength
$\lambda_c$ of order $\sim 1-10 \ {\rm \mu m}$ (theory shows that shorter wavelength roughness
is unimportant), and gives friction coefficients of order unity, 
as indeed observed experimentally.

\begin{figure}[htb]
  \includegraphics[width=0.30\textwidth]{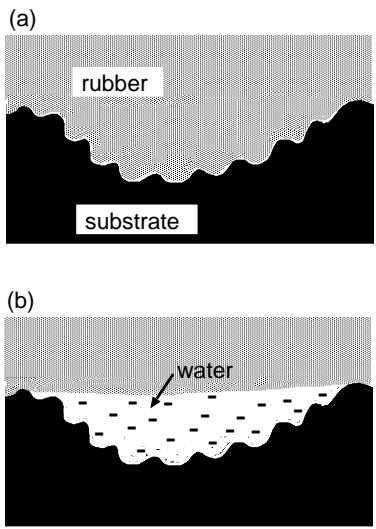}
  \caption{
  \label{sealing}
A rubber block sliding on a rough hard substrate. (a) The rubber penetrates into a 
large substrate valley and explores the short
wavelength roughness in the valley. The pulsating rubber deformations 
induced by the short-wavelength roughness contribute to 
the friction force. 
(b) On a wet substrate the water trapped in the large valley 
forms a pool preventing the rubber from
penetrating into the valley. It will hence
remove the valley contribution to the friction force 
This rubber {\it sealing effect} reduces the sliding
friction.
    }
\end{figure}

In this paper we study rubber friction at low sliding velocities
on wet rough substrates, where it has been observed that the 
friction typically is $20-30\%$ smaller than
for the corresponding dry surfaces, 
see Fig.~\ref{measured.mu.slip} and Ref.~\cite{Gert,MeyerWalter}. 
We show that this cannot be a hydrodynamic
effect,(see especially Appendix A).
Expanding on our recent
proposal,\cite{nmat} we put forward a novel explanation 
based on the rubber sealing off
pools, namely regions on the substrate filled with water as shown in Fig.~\ref{sealing}. 
The water effectively smoothens the substrate surface, and thus reduces 
the viscoelastic deformation contribution to the rubber friction from
the surface asperities.

\vskip 0.3cm

{\bf 2. Theory}

\begin{figure}[htb]
  \includegraphics[width=0.35\textwidth]{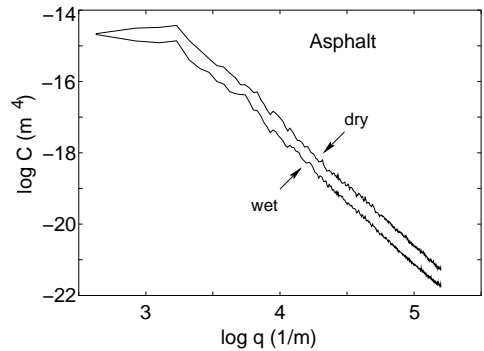}
  \caption{
  \label{Cq}
The logarithm (to base 10) of the surface roughness power spectra $C(q)$ for
a dry and a wet asphalt road surface, as a function of the logarithm of
the wavevector $q$. 
    }
\end{figure}

The contribution to rubber friction from the viscoelastic deformation of the rubber
surface by the substrate asperities depends only on the complex frequency-dependent
viscoelastic modulus $E(\omega )$ of the rubber and on the substrate surface roughness
power spectrum $C(q)$, which is defined as follows. Let $h({\bf x})$ be the substrate height
profile measured from the average 
surface plane defined so that $\langle h \rangle = 0$,
where $\langle ... \rangle$ stands for ensemble averaging or averaging over the
total surface. We then have
$$C(q) = {1\over (2\pi )^2}
\int d^2 x \ \langle h({\bf x}) h({\bf 0}) \rangle e^{-i{\bf q}\cdot {\bf x}}$$
We assume that the statistical properties of the substrate surface are isotropic
and translationally invariant 
(within the surface plane), 
so that $C(q)$ only depends on the
magnitude $q=|{\bf q}|$ of the wave vector ${\bf q}$. The upper curve in Fig.~\ref{Cq} shows the
power spectrum calculated from the height profile $h({\bf x})$ measured for an asphalt road
using an optical method.
The figure shows $C(q)$ as a function of $q$ on a log-log scale.
For $q > 1600 \ {\rm m}^{-1}$, $C(q)$ shows a power law dependence on the wave vector $q$, as 
expected for a self affine fractal surface. The fractal dimension of the surface is about
2.2 and the root-mean-square roughness $h_{\rm rms} \approx 0.3 \ {\rm mm}$. 
For $q < q_0$, $C(q)$ is constant. The roll-off wavevector $q_0$
correspond to the wavelength $\lambda_0 = 2\pi /q_0 \approx 4 \ {\rm mm}$ and reflect the
largest asperities or sand particles contained in the asphalt.

In general, the hysteresis contribution to rubber friction increases with increasing 
magnitude of $C(q)$. However, the friction depends on $C(q)$ over a wide range of
wave vectors $q$. For example, the rubber friction on 
asphalt road surfaces depends on $C(q)$ for $q_0 < q < q_1$, where typically $q_0\approx 10^3
 \ {\rm m^{-1}}$ and $q_1 \approx
10^6 \ {\rm m}^{-1}$. 
For a wet road surface, the rubber
will seal some surface areas filled with water (pools) as schematically
shown in Fig.~\ref{sealing},
and this leads to an effective smoothening of the substrate and to a reduced
power spectrum. We illustrate this below for the same asphalt road surface for which we
showed the power spectrum in Fig.~\ref{Cq}(top curve). 

Consider a tire rolling or sliding on a wet road surface. In the Appendix A we show
in detail that at low velocities (say $v < 30 \ {\rm km/h}$), there is 
a negligible hydrodynamic water build up between the tire and the road
surface. There is sufficient time for the water to be squeezed 
from the contact regions between the tire and the road surface, 
{\it except} for water trapped in road cavities and sealed
off by the road-rubber contact at the upper boundaries of the cavities (see Fig.~\ref{sealing}).
Thus, in what follows we will only focus on the smoothing effect on the road profile
by the sealed off water 
pools.

\begin{figure}[htb]
  \includegraphics[width=0.35\textwidth]{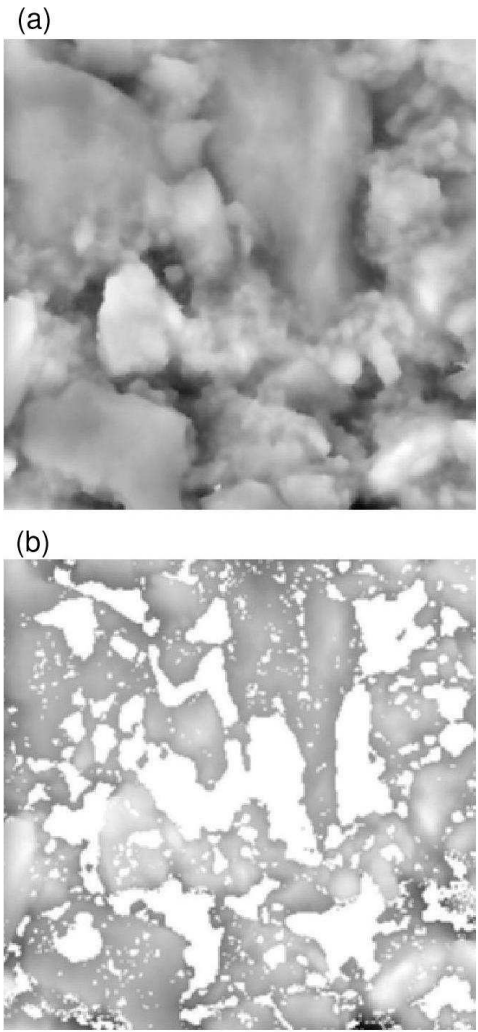}
  \caption{
  \label{topo}
Optically observed contour line height profile of (a) a dry asphalt road ($1.5 \ {\rm cm} 
\times 1.5 \ {\rm cm}$ area)
and (b) the calculated profile for the same surface area when wet.    
Deeper asphalt regions are darker, and the water pools in (b) are white.
}
\end{figure}

In Fig.~\ref{topo}(a) we show the height contour 
lines of a square $1.5 \ {\rm cm}\times
1.5 \ {\rm cm}$ area of the dry asphalt road.
We have calculated the height profile $h'$ shown in Fig.~\ref{topo}(b) (wet surface)
numerically. Every valley has been filled with water up to the maximum
level where the water still remains confined, i.e., up to the lowest
point of the edge surrounding the pool. Any extra addition of water
would flow out of the square area. This criterion to fill the
surface with water is shown schematically in Fig.~\ref{sealing}.
 From the new height profile $h'({\bf x})$ we can calculate a new power
spectrum $C'(q)$ shown by the
lower curve in Fig.~\ref{Cq}. Now we make the basic assumption 
(see also below) that when a rubber block
slides on the wet surface, the friction force will be determined 
by the power spectrum $C'(q)$.
This implies that the water in the pools is sealed off by 
the rubber (as indicated in Fig.~\ref{sealing})
and cannot get squeezed out. This prevents the rubber 
from penetrating into the corresponding valley, 
and will reduce the sliding friction by removing contact 
with the rough walls of the valley.  
 
We notice that our filling criterion is generally not unique, since it
depends on the size of the surface area we are considering. In fact it
becomes unique (apart from small differences localized at the borders)
when the size is much larger than the roll-off wavelength
$\lambda_0\approx 4$ mm, which corresponds to the typical size of the
largest pools.  Nonetheless a realistic description requires the surface
area to be comparable with the size of the tread block of the tire,
while water at the boundaries does not get trapped but it is free to
flow away across the channels of the tread pattern. This is indeed the
conditions we are adopting through the choice of the size and the
boundary conditions of the filling procedure. In Appendix B we present
the results for another asphalt surface, confirming our results despite
the unavoidable statistical noise.

\vskip 0.3cm

{\bf 3. Numerical results}

Now we present numerical results related to tire friction on dry and wet
substrates, calculated using the theory presented in 
Ref.~\cite{PerssonJCP,PerssonTBP}. Neglecting the flash temperature, the friction coefficient 
is given by\cite{PerssonJCP}
$$\mu = {1\over 2}\int dq \ q^3 C(q) P(q)
\int_0^{2\pi} d\phi \ {\rm cos} \phi \ {\rm Im}{E(qv \ {\rm cos}\phi ) \over
(1-\nu^2)\sigma}$$
where 
$$P(q)= {2\over \pi} \int_0^\infty dx \ {{\rm sin}x \over x} 
{\rm exp}\left [-x^2G(q) \right ]= {\rm erf}\left(1/2\surd G\right )$$
with
$$G(q)={1\over 8}\int_0^q dq \ q^3C(q)\int_0^{2\pi}d\phi \ \left |{E(qv \ {\rm cos}\phi ) \over
(1-\nu^2)\sigma} \right |^2$$
where $\sigma$ is the perpendicular pressure (the load divided by the nominal contact area).
 
The results presented below
have been obtained for a standard 
tread compound, sliding on the asphalt road introduced in Sect. 2. 
We use the measured
complex viscoelastic modulus of the rubber
and the power spectra presented in
Fig.~\ref{Cq} for the dry and wet road surfaces.
 
\begin{figure}[htb]
  \includegraphics[width=0.35\textwidth]{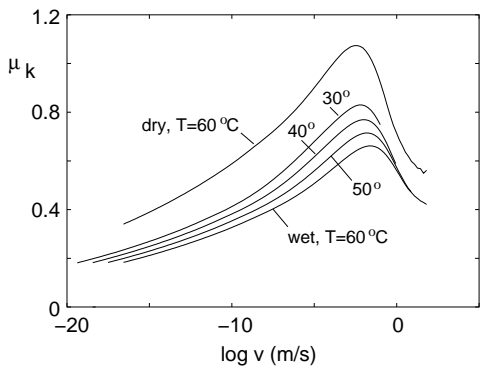}
  \caption{
  \label{mukinetic}
Kinetic friction coefficient as a function of the logarithm of
the sliding velocity, calculated
for a standard
tread compound and an asphalt substrate
    }
\end{figure}

In Fig.~\ref{mukinetic} we show the kinetic friction coefficient 
calculated 
for the dry surface at $T=60^\circ \ {\rm C}$ (a typical 
tire temperature during driving on a dry road),
and for the wet surface at four different temperatures, namely $T=30$, $40$, $50$, and
$60^\circ \ {\rm C}$. Note that
on a wet road the tire temperature is generally lower than on 
the dry surface, its typical value being $\sim 30^\circ \ {\rm C}$. 
The decreased friction with increasing temperature shown in Fig.~\ref{mukinetic}
is always observed for rubber, and 
results from the shift in the 
viscoelastic spectrum to higher frequencies with increasing
temperature (temperature makes rubber more 
elastic and less viscous), which in turn 
reduces rubber friction. 

\begin{figure}[htb]
  \includegraphics[width=0.35\textwidth]{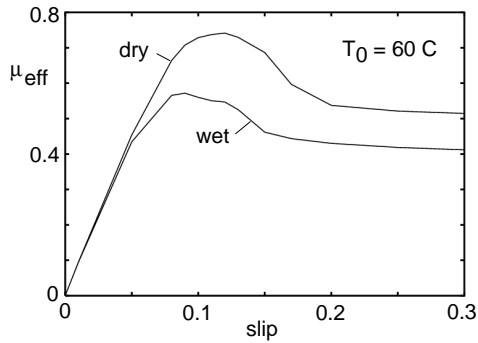}
  \caption{    \label{muslip}
The effective friction coefficient as a function 
of slip for dry and wet road surface,
calculated
for a standard
tread compound and an asphalt substrate.
    }
\end{figure}

All modern cars use Anti-Blocking Systems (ABS). In this case during 
braking the wheels never get fully locked, but the rolling velocity, 
$v_R=\omega R$, is smaller than the forward velocity $v$ of the car,
implying that some slip must occur at the tire-road interface. 
The fundamental characteristic
of the tire-road friction relevant for ABS braking is the so called $\mu$-slip curve.
Here the slip is defined as $s=(v-v_R)/v$. Hence $s=0$ correspond to pure rolling, and
$s=1$ to locked wheel braking. The $\mu$-slip curve depends not only on the rubber-road friction but also
on the elastic properties of the tire. Thus, at small slip the tire tread blocks are not slipping
relative to the road surface as they first enter the foot print contact area, but will only slip
close to the exit of the foot print contact area. 
The theoretically calculated $\mu$-slip curves 
(see Fig.~\ref{muslip})
for dry and wet surfaces are similar to experimental results (Fig.~\ref{measured.mu.slip}).
In particular, the $\mu$-slip curve for the dry 
and wet surface exhibit a similar dependence
on the slip
as is also observed experimentally\cite{Gert,MeyerWalter}, 
but would 
not be expected if hydrodynamic effects  
were the origin of the decrease in $\mu_{\rm eff}$ for wet surfaces. 
In that case one would
expect a much stronger reduction of $\mu_{\rm eff}$ for large slip.

In Appendix B we present numerical results 
for another asphalt surface with nearly twice
as large surface {\it rms} roughness amplitude. Nevertheless, the difference between
the friction coefficients for the 
dry and wet road surfaces is very similar to what we have found above. This shows that
the conclusions above are of general validity. 

Finally, all sealings leak. This is in particularly true in the present case
because the upper boundary 
of a water filled cavity,
which is in contact with the rubber,
is not smooth, but
has roughness on many length scales, and one cannot expect
the rubber to make perfect contact with this region of the substrate. 
Thus, one expects narrow channels through which
water may leak out. As a result, for {\it very low} car 
velocities the water may have a
negligible influence on the rubber friction. In fact, experiments
have shown that the difference in $\mu_{\rm eff}$
between dry and wet surfaces for velocities $v < 1 \ {\rm m/s}$ is very small.

\vskip 0.3cm

{\bf 4. Discussion}

We have shown that the reduction in rubber friction usually 
observed when a hard rough road surface becomes wet with water
cannot be explained as a hydrodynamic effect, and 
we have proposed a
novel mechanism involving the rubber sealing off pools filled with water.
This leads to an effective smoothening of the substrate 
and to a lower sliding friction. However, the 
sealing mechanism 
may be more complex than outlined above. One can 
imagine dynamical processes where sealing can occur
although not allowed by our procedure. 
In fact, long-lived trapped water regions have been observed 
even when a rubber ball is squeezed against a
smooth flat substrate\cite{Roberts,Mugele}. The trapped 
water regions sometimes exist even
after several hours of stationary contact, therefore
reducing the friction on wet surfaces. Another complication is 
that 
the water is often located in ``deep'' valleys which
contribute little to the sliding friction, since the rubber is 
not able to deform enough to fill them out. Hence, our 
calculation tend to overestimate the
influence of such deep valleys to the change in the rubber friction.  
Since the calculated difference between the friction on
dry and wet 
surfaces is of similar magnitude to that 
observed,  
we suggest that the above two effects  
tend to cancel each other. 

Another effect which has been suggested to influence rubber 
friction on wet surfaces is the dewetting transition\cite{wet1,wet2},
which has been studied mainly for very smooth surfaces.
The stability of a water film between a rubber block and a flat 
solid substrate is controlled by
the spreading parameter:
$$\Delta \gamma = \gamma_{\rm RS}- (\gamma_{\rm RL}+\gamma_{\rm LS})$$
where $\gamma_{\rm RS}$, $\gamma_{\rm RL}$ and $\gamma_{\rm LS}$ are the rubber/solid,
rubber/liquid and liquid/solid interfacial free energies per unit area. If $\Delta \gamma >
0$ the liquid film (in the absence of a squeezing force) is stable. If $\Delta \gamma < 0$
the flat liquid film is unstable and is expected to dewet by 
nucleation and growth of a dry patch
surrounded by a rim, collecting the rejected liquid. 
However, we do not believe that the dewetting transition is crucial
in the context of (rough surface) tire-road friction.
First, the dry state should not be
the minimum free energy state, since water wets rock surfaces, 
(which usually consist of polar oxides), and this
should favor a state with an intercalated
water film between the surfaces. Second, the dewetting 
transition usually involves a thermally activated nucleation 
process. Thus it should have a strong dependence of temperature,
while such strong dependence is not observed for
the friction force.
Third, the dewetting transition is unlikely to affect the water 
sealed off by the rubber. Finally, we have argued that the 
adhesive interaction gives a negligible contribution to the 
rubber friction force on very rough surfaces so that it is
irrelevant whether or not a very thin water film (thickness $h < 1 \ {\rm \mu m}$) 
is present at the rubber-road asperity
contact areas.

\vskip 0.3cm

{\bf 5. Summary and conclusion}

Rubber friction on wet rough substrates at low velocities is 
typically $20-30\%$ smaller than for the corresponding dry surfaces. 
We have shown that this cannot be due 
to hydrodynamics, and we have propose a novel explanation based on a sealing 
effect exerted by rubber on substrate ``pools'' filled with water. 
Water effectively smoothens the substrate, reducing the major 
friction contribution due to induced viscoelastic deformations 
of the rubber by surface asperities. The theory was illustrated with 
applications related to tire-road friction.

\vskip 0.3cm

{\bf Acknowledgments}
Work in SISSA was sponsored by the Italian Ministry of University and
Research through MIUR COFIN 2003, MIUR COFIN 2004 and MIUR FIRB RBAU01LX5H as
well as through Istituto Nazionale Fisica della Materia, grant INFM FIRB
RBAU017S8R. We acknowledge computational support from CINECA
(Casalecchio, Bologna).

\vskip 1cm

{\bf APPENDIX A: Hydrodynamic squeeze-out}

Here we present simple arguments to demonstrate that 
there is negligible hydrodynamic 
water film build up 
at low car velocities,
between the road surface and the tread blocks, which
is a necessary condition for the sealing mechanism to be relevant.
We are interested in water squeeze-out from the rubber-road asperity contact areas,
down to a thickness of order $h_c$, where 
$h_c=h_{\rm rms}(\lambda_c)$ is
the surface root-mean-square roughness amplitude 
derived from surface roughness wavelength components larger than
$\lambda_c$. This is the shortest surface roughness component which effectively contribute to
the rubber friction on the dry surface
(typically $\lambda_c \approx 5 \ {\rm \mu m}$ and $h_c\approx
2 \ {\rm \mu m}$). We first study the squeeze-out on a length scale larger than the
road rms roughness, which typically is of order 1 mm or less. In this case we can
neglect the surface roughness and assume that the road surface is completely flat.
We consider two limiting cases, namely a viscous liquid without inertia effects and 
a liquid with inertia but neglecting the viscosity.

\vskip 0.3cm

{\bf Role of viscosity}

\begin{figure}[htb]
  \includegraphics[width=0.35\textwidth]{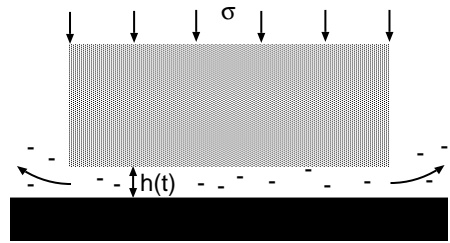}
  \caption{
  \label{squeeze}
A tread block squeezed against a smooth flat substrate in a liquid. The surface
separation $h(t)$ decreases with increasing time.    
}
\end{figure}

Consider first the influence of the water viscosity on the squeeze-out of the water between a 
tire tread block and the substrate. 
We assume first that the substrate is perfectly flat and we neglect the
deformation of the tread block, i.e. the bottom surface of the tread block is 
considered flat (see Fig.~\ref{squeeze}). 
If the tread block is squeezed with the stress $\sigma$ 
against the substrate in water, and if the
thickness of the water layer is $h_0$ at time $t=0$, then 
(neglecting inertia effects) 
the thickness $h=h(t)$ at time $t$
is given by\cite{Persson1} 
$${1\over h^2(t)} - {1\over h_0^2} = {16 t \sigma \over 3 \mu D^2},\eqno(1)$$
where $\mu$ is the viscosity and $D$ the width of the tread block. During pure rolling,
or rolling--sliding with small slip,
the tread block stays a time $t \approx W/v$ in the tire foot print area,
where $W$ is the length of the foot print contact area, and $v$ is the 
tire rolling velocity. 
Since we are interested in 
$h(t)\ll h_0$, it follows from (1) that the thickness $h_1$ of the water film at time 
$t=W/v$ satisfies: 
$${1\over h_1^2} \approx {16 W \sigma \over 3 v \mu D^2}$$
or
$$v \approx 
{16 W h_1^2 \sigma \over 3 \mu D^2}\eqno(2)$$
If we take $h_1 = 1 \ {\rm mm}$, the tread block diameter $D=3 \ {\rm cm}$, the
foot print length $W=10 \ {\rm cm}$ and contact pressure $\sigma = 1 \ {\rm MPa}$ we get
for water ($\mu \approx 10^{-3}\ {\rm Ns/m^2} $):
$v\approx 10^6 \ {\rm m/s}$. Thus, the the viscosity of the water is irrelevant
for the initial squeeze-out down to a thickness of order the root-mean-square
amplitude of the substrate roughness.

If we consider pure sliding, the relation between the sliding velocity
and the shortest separation between the 
tread block and the substrate will be\cite{Persson2}
$$v \approx 
{h_1^2 \sigma \over \alpha  \mu D},\eqno(3)$$
where $\alpha$ depends on the ratio of the tread-block substrate separation at the inlet and
the exit of the junction. Typically $\alpha \approx 0.1$. 
Thus, 
to within a factor of order unity, 
(3) can be obtained from (2) if we put $W=D$, 
and the estimate of $v$ given above still holds. 

\vskip 0.3cm

{\bf Elastohydrodynamic}

The analysis above has assumed a flat substrate. 
However, a road surface has a surface roughness
with a typical root-mean-square amplitude of about 1 mm,
and the analysis above can only be applied until
$h(t)\approx 1 \ {\rm mm}$. In studying the influence of 
surface roughness on the squeeze-out, we consider first the longest-wavelength
roughness, with a wavelength determined by the roll-off wavevector
$q_0$ via $\lambda_0 = 2 \pi /q_0$. When the system is studied at the lateral resolution
$\lambda_0$, 
the contact between the rubber and the substrate occurs at randomly distributed 
asperities with the radius of curvature $R \approx (h_{\rm rms}q_0^2)^{-1}$. We denote these
asperities as macro asperities because they are the largest asperities occurring on the
substrate.

\begin{figure}[htb]
  \includegraphics[width=0.35\textwidth]{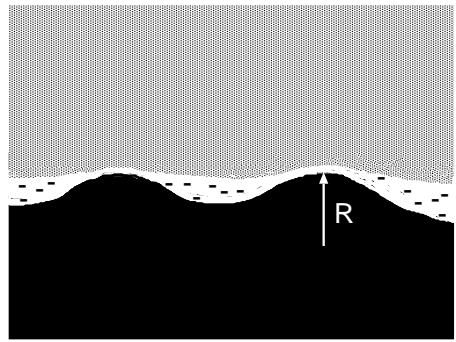}
  \caption{
  \label{squeeze.bump}
A tread block squeezed against a rough flat substrate in a liquid. 
}
\end{figure}

Consider a tread block squeezed against a road macro asperity
in water (see Fig.~\ref{squeeze.bump})  and assume 
that the squeezing force equals $F$, and that the rubber slides with the velocity $v$
relative to the asperity. The thickness of the water layer between the asperity and the rubber surface 
can be estimated using the following standard results from elastohydrodynamics\cite{Hamrock}:
$$v \approx {0.16\over \mu}\left ({(E^*)^9F^4h_1^{20}\over R^{15}}\right )^{1\over 13}$$ 
When the rubber-substrate interface is studied with a lateral resolution of order $\lambda_0$,
the area of contact is
about $10 \% $ of the nominal contact area, and the loading force on a macro asperity 
will typically be
$F=100 \ {\rm N}$.
Using $R=2 \ {\rm mm}$, 
$E^* = 1 \ {\rm MPa}$ 
and $h_1 = h_c \approx 1 \ {\rm \mu m}$ 
it follows $v\approx 200 \ {\rm m/s}$.
In this study we have neglected the (short-wavelength)
roughness on the macro asperity. However, neglecting sealing of water pools,
it is easy to see
that the inclusion of the short-wavelength roughness in the range 
$\lambda_c < \lambda < \lambda_0$
can only facilitate (speed up) the squeeze-out
of the water, 
down to the water thickness $\sim h_c$  
at the micro-asperities, characterized by the wavelength 
$\lambda_c$. This result follows from the fact that  
the average space between the surfaces at the macro asperity will
be much larger than $h_c$.

\vskip 0.3cm

{\bf Role of inertia}

Let us now study the influence of the inertia of the water on the squeeze out
at a tread block. Neglecting the viscosity, the pressure work (per unit time)  
$-\sigma D^2 \dot h$ must be equal to the change in the water kinetic energy
per unit time, $\dot K$. The kinetic energy is of order
$$K \approx \rho  D^2 h \bar v^2,$$
where the average velocity
$\bar v \approx D\dot h / h$. Thus we get [with $h(0)=h_0$]:
$$-\sigma  [h(t)-h_0] \approx \rho D^2 \dot h^2/h$$
or
$$- \sigma [h(t)-h_0] h(t) \approx \rho D^2 \dot h^2$$
We are interested in the case $h(t) \ll h_0$ so we can approximate
$$\sigma h_0 h(t) \approx \rho D^2 \dot h^2,$$
which gives the squeeze-out time (i.e., $h(t)=0$):
$$t \approx D \left ({\rho \over \sigma}\right )^{1/2}\eqno(4)$$
The time the tread block spends in the foot print area is (for small slip)
of order $W/v$ so that (4) gives
$$v \approx {W\over D} \left ({\sigma\over \rho}\right )^{1/2}\eqno(5)$$
Using the same numerical values for $W$, $D$ and $\sigma$ as before gives for
water ($\rho = 10^3 \ {\rm kg/m^3}$): $v \approx 100 \ {\rm m/s}$. Thus, 
if $v \ll 100 \ {\rm m/s}$ the inertia of the water 
will not inhibit the water squeezed out from the interface.
When the viscosity is neglected, the total squeeze-out time is finite,
but complete squeeze-out 
(within the framework of the Navier Stokes equations) takes
infinitely long time (see Eq. (1)). 
As a result 
the viscosity effect
will always dominate over the inertia effect
for very thin liquid films, 
and inertia can be neglected. 
However, as shown above, for water film thickness $h > 1 \ {\rm \mu m}$ this is not the case
and the water viscosity can be neglected.

\vskip 0.3cm

{\bf Aquaplaning (hydroplaning)}

Aquaplaning (or hydroplaning) 
refer to the limiting case when a tire is completely separated from the road surface
by a liquid film. Here we will only consider a tire 
without a tread pattern. 
In the case of clean water, aquaplaning is entirely due to the inertia of the water,
and viscous effects are negligible\cite{Andren}. 
This can be seen by applying eq. (3) with
$D=W\approx 10 \ {\rm cm}$ equal to the length of the footprint area 
and $h_1\approx 1 \ {\rm mm}$ equal to the amplitude of the
road surface roughness.
This gives $v\approx 10^5 \ {\rm m/s}$, i.e., larger than
observed experimentally by a factor of $10^4$. 
On the other hand, the inertia effect is important even 
at relatively low velocities. Thus, from (5) (with $W=D$)
$$v \approx \left ({\sigma\over \rho}\right )^{1/2},$$
we get $v\approx 20 \ {\rm m/s}$. In fact, some tire road lack of contact will occur at the front of the
footprint contact 
area already at lower sliding velocity, but an accurate study of this effect requires
taking into account the deformations of the tire, and is possible only using 
advanced finite element calculations. 

\begin{figure}[htb]
  \includegraphics[width=0.35\textwidth]{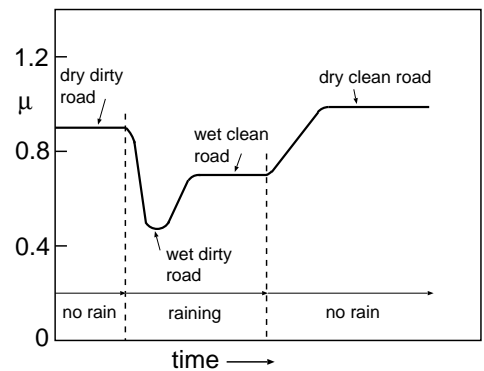}
  \caption{
  \label{mutime}
The dependence of the tire-road friction coefficient on time during rain.
}
\end{figure}

Viscous effects may also be important for aquaplaning 
if the road surface is covered by a high viscosity fluid, e.g., oil
spill or mud, since these fluids may have viscosities 
$\sim 1000$ (or more) times higher than that of water.
Many drivers will have noticed that roads are sometimes 
most slippery when rain begins, and this is caused by rain mixing
with road debris, such as dirt (e.g., stone particles or rubber wear particles) and oil, creating an
effective high viscosity lubricant (similar to clay mixed with water)
that will decrease the coefficient of friction (see Fig.~\ref{mutime}).
The coefficient of friction will be particularly low after long 
time periods, due to the build 
up of road debris. As Fig.~\ref{mutime} shows, the coefficient of friction between the road surface
and the tire will increase as the rain washes away the road debris. The maximum friction will result
when the road has dried, as it is now free from particle contamination (the particles have been washed
away by the rain).

To summarize, we have shown that the water viscosity is irrelevant for squeeze-out 
(unless the effective viscosity
is strongly enhanced by contamination), while the water inertia will be important 
for sufficiently high sliding
or rolling velocities. However, for thin water films (less than the tread height), where aquaplaning
will not occur, for velocities below, say $\sim 30 \ {\rm km/h}$, the water
inertia effect can also be neglected,
and the only way the water will affect the rubber friction is via the sealing effect.  

\vskip 1cm

{\bf APPENDIX B: results for another asphalt surface}

\begin{figure}[htb]
  \includegraphics[width=0.35\textwidth]{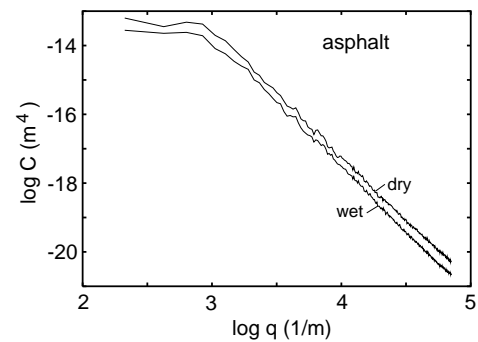}
  \caption{
  \label{Cq2}
The logarithm of the surface roughness power spectra $C(q)$ for a dry 
and a wet asphalt road surface, as a function of the logarithm of
the wavevector $q$. 
    }
\end{figure}

\begin{figure}[htb]
  \includegraphics[width=0.35\textwidth]{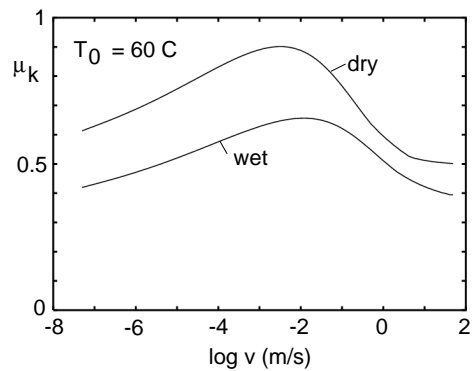}
  \caption{
  \label{mukinetic2}
Kinetic friction coefficient as a function of the logarithm of
the sliding velocity, calculated
for a standard
tread compound and an asphalt substrate.
    }
\end{figure}

To demonstrate the general nature of the results presented above, here we present
results for a second asphalt road with nearly twice as large {\it rms} roughness amplitude as
the for asphalt surface used above. In Fig.~\ref{Cq2} we show the power spectra for the dry and 
wet asphalt surfaces. Fig.~\ref{mukinetic2} shows the kinetic friction coefficient as a function
of the logarithm of the sliding velocity both for the dry and wet surfaces. 

Fig.~\ref{muslip} shows the effective friction coefficient as a function of the slip. The 
figure is obtained from a computer simulation, 
where the motion of a single tread block 
in the tire-road footprint contact area is studied.
However, a more realistic calculation involving all the tread blocks coupled to each other
(indirectly) via the car cass elasticity, should give a similar result.
The $\mu$-slip curves presented in the figure are in good qualitative agreement with typical measured
$\mu$-slip curves, and shows a similar reduction in the friction as the kinetic friction coefficients
shown in Fig.~\ref{mukinetic2}.

\end{document}